\documentclass[12pt]{article}
\usepackage{graphicx}
\usepackage{cite}
\usepackage[latin1]{inputenc}
\let\section=\subsection     \let\subsection=\subsubsection                
\renewcommand\thesubsection{\arabic{subsection}}                           

\begin{document}
\begin{center}
   {\large \bf LIQUID-GAS COEXISTENCE REGION IN CENTRAL Xe+Sn REACTIONS}\\[5mm]
    \underline{B.~BORDERIE}$^1$, M.F RIVET$^1$, G.~T\u{A}B\u{A}CARU$^{1,2}$,
   M.~COLONNA$^4$, P.~D\'ESESQUELLES$^9$, M.~P\^ARLOG$^2$,   \\
   and  \\
 G.~AUGER$^3$,  Ch. O. BACRI$^1$, N.~BELLAIZE$^5$, R.~BOUGAULT$^5$,
 B.~BOURIQUET$^3$, A.~BUTA$^5$,A.~CHBIHI$^3$,  J.~COLIN$^5$,
 A.~DEMEYER$^7$, E.~GALICHET$^{1,10}$, E.~GERLIC$^7$, D.~GUINET$^7$,
 B.~GUIOT$^3$, S.~HUDAN$^3$,P.~LAUTESSE$^7$,
 F. LAVAUD$^1$, J.L.~LAVILLE$^3$, J.F.~LECOLLEY$^5$, C.~LEDUC$^7$
 R.~LEGRAIN$^6$, O.~LOPEZ$^5$, M.~LOUVEL$^5$,
 J.~NORMAND$^5$, P.~PAWLOWSKI$^1$,
 E.~ROSATO$^8$,
 J.C.~STECKMEYER$^5$, B.~TAMAIN$^5$,
 L.~TASSAN-GOT$^1$, E.~VIENT$^5$, J.P.~WIELECZKO$^3$ \\[2mm]
  INDRA Collaboration \\[5mm]
{\small\textit{
$^1$ Institut de Physique Nucl\'eaire, IN2P3-CNRS, F-91406 Orsay Cedex,
France.~\\
$^2$ National Institute for Physics and Nuclear Engineering, RO-76900
Bucharest-M\u{a}gurele, Romania.~\\
$^3$ GANIL, CEA et IN2P3-CNRS, B.P.~5027, F-14076 Caen Cedex, France.~\\
$^4$ Laboratorio Nazionale del Sud, Viale Andrea Doria, I-95129 Catania,
Italy.~\\
$^5$ LPC, IN2P3-CNRS, ISMRA et Universit\'e, F-14050 Caen Cedex, France.~\\
$^6$ DAPNIA/SPhN, CEA/Saclay, F-91191 Gif sur Yvette Cedex, France.~\\
$^7$ Institut de Physique Nucl\'eaire, IN2P3-CNRS et Universit\'e, F-69622
Villeurbanne Cedex, France.\\
$^8$ Dipartimento di Scienze Fisiche e Sezione INFN, Università di Napoli\\
``Federico II'', I80126 Napoli, Italy.~\\
$^9$ Institut des Sciences Nucl\'eaires, IN2P3-CNRS et Universit\'e, F-38026\\
Grenoble Cedex, France.~\\
$^{10}$ Conservatoire National des Arts et M\'etiers, F-75141 Paris cedex
03.~\\[8mm]}}
\end{center}

\begin{abstract}\noindent
     Charge partitions and distributions of fragments emitted
     in multifragmentation of fused
     systems produced in central collisions are studied over the incident
     energy range 32-50 MeV per nucleon. Most of the charged products are
     well identified thanks to the high performances of the INDRA 4$\pi$
     array. Supported by dynamical calculations, charge correlations are used to
     evidence, or not, spinodal instabilities and consequently the liquid-gas
     coexistence region over the considered incident energy
     range.  It was claimed
     in the last few years that mass/charge distributions should follow a 
     power law behavior
      in the coexistence region. The Z distributions measured are discussed.
     A first attempt is made to derive in which Z region the border between
     liquid and gas parts is located.
\end{abstract}

\section{Introduction}

Despite the large amount of experimental studies regarding the nuclear
liquid-gas phase transition, fundamental questions have mostly eluded conclusive
answers. We shall discuss two questions in this contribution. A first one
regards the mechanism of phase separation for
matter which is expanding through a density-temperature region inside the
liquid-gas coexistence region. And the second is whether power laws claimed
to be observed for fragment mass/charge distributions are related or not
to this region;
power laws are indeed only expected at or near the critical point.

We report here on studies performed with INDRA~\cite{indra} of
multifragmentation of very
heavy fused systems formed in central collisions between
 $^{129}Xe$ and $^{nat}Sn$ at 32,39,45 and 50 AMeV.
 These fused systems can
be identified to well defined pieces of nuclear matter and eventually
 reveal fragmentation properties to be compared to models in which bulk
 instabilities are present.

Detailed information on the experiment, on calibration, identification and
on the selection of fused events
can be found in~\cite{MA97,RI98,FR001,TA99,PA00}. Reaction products with
charge Z$\geq$5 were defined as fragments.


\section{Enhancement of equal-sized fragment partitions and spinodal
instabilities }

\begin{figure}
\includegraphics[width= 14cm]
{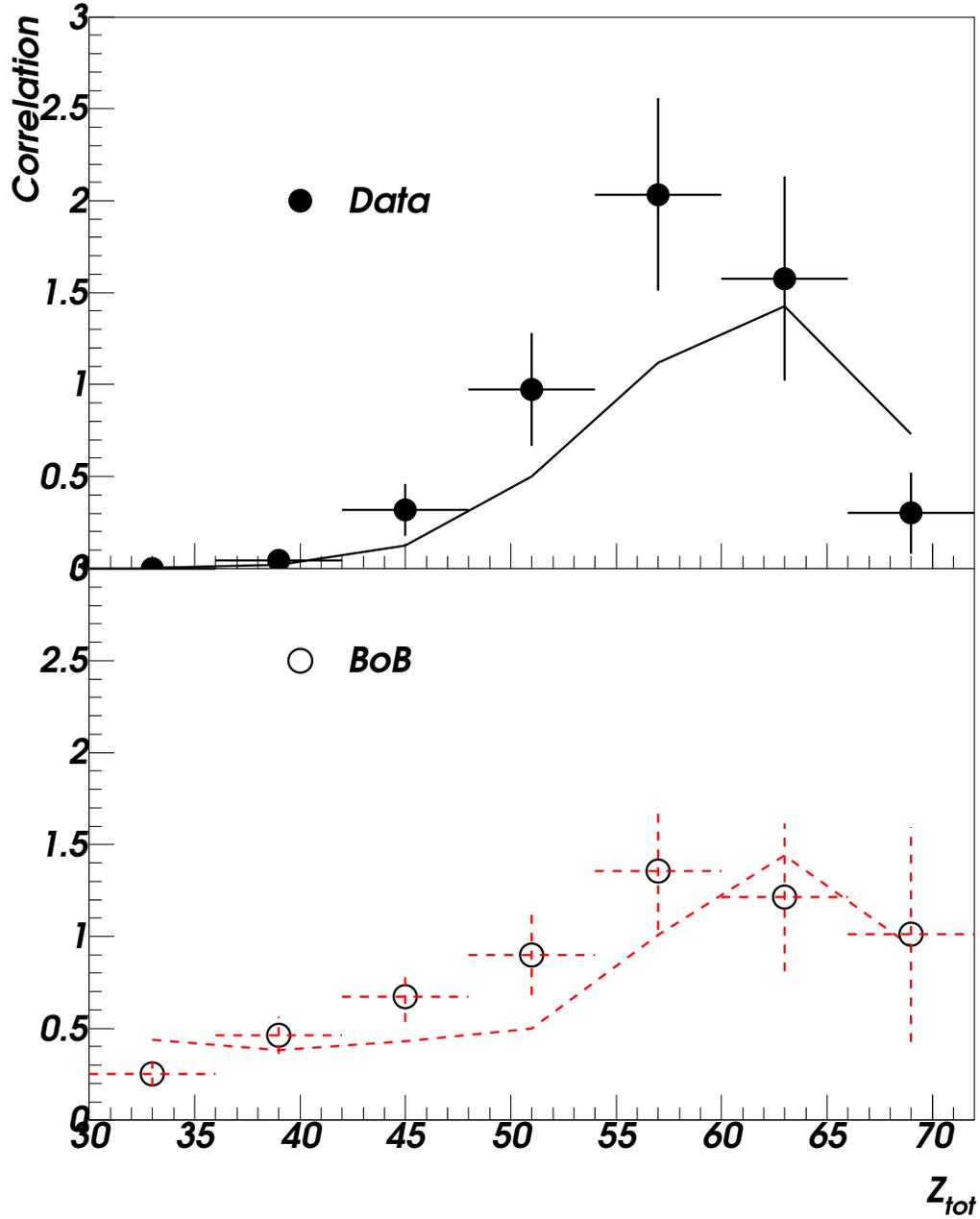}
\caption{Higher-order charge correlations : quantitative results
for experimental data (up) and simulations (down).
Symbols indicate the events where $\Delta$Z = 0-1, curves show
the background (see text).
Vertical bars correspond to statistical errors and horizontal
bars define $Z_{tot}$ bins.}
\label{cor_exp_quant}
\end{figure}

Many theories have been developed to explain
multifragmentation (see for example ref.~\cite{MO93} for a general
review of models).
One can come in particular to the concept of multifragmentation by
considering volume instabilities of the spinodal type.
Indeed during a collision, a wide zone of the nuclear
matter phase diagram may be explored and the nuclear system may enter
the liquid-gas phase coexistence region (at low density) and even more
precisely the unstable
spinodal region (domain of negative compressibility). Thus, a possible
origin of multifragmentation may be found through the growth of density
fluctuations in this unstable region. Within this theoretical scenario a
breakup into nearly equal-sized ``primitive'' fragments should be favored
in relation with the wave-lengths of the most unstable modes present in the
spinodal region~\cite{AY95}. However this simple picture is expected to be
strongly blurred by several effects: the beating of different modes,
the presence of large wavelength instabilities, eventual coalescence
of nascent fragments, secondary decay of excited fragments and mainly the
finite size of the system~\cite{JA96,CO97}. Therefore only a weak proportion
of multifragmentation events with nearly equal-sized fragments is expected.
To search for such events a very sensitive correlation method was used,
which is called
``higher order correlations'' and was proposed in ref~\cite{MO96}.
Its originality consists in the fact that all information on fragments of one
event is condensed in two variables (average fragment charge $<Z>$ and
standard deviation $\triangle$Z).

The charge correlation is defined by the expression:

\begin{equation}
\left. \frac{Y(\Delta Z, <Z>)}{Y'(\Delta Z, <Z>)} \right| _{M} 
\end{equation}

Here, $Y(\Delta Z, <Z>)$ is the yield of selected events with
$<Z>$ and $\Delta Z$ values and
$M$ is the fragment multiplicity.
%
%

The denominator $Y'(\Delta Z, <Z>)$ which represents the  
uncorrelated yield is built, for each fragment multiplicity, by taking 
fragments in different events of the selected sample. The number of
uncorrelated events was
chosen large enough (10$^3$ per true event) to strongly reduce 
their contribution to statistical error. 
With such a correlation method, if events with  nearly equal-sized fragments
are produced, we expect to see peaks appearing at $\Delta$Z values close to
zero. Taking into account secondary decay of fragments, the bin $\Delta$Z=0-1
was only considered. At 32 AMeV incident energy peaks were observed in this
bin for each fragment multiplicity~\cite{TA00,BO01}.

%

We can now estimate whether the enhancement of events with
equal-sized fragments is statistically significant and quantify their
occurrence. In this aim
we built charge correlations for all events, whatever
their multiplicity, by replacing the variable $<Z>$ by
$Z_{tot}$ = M$\times <Z>$. For this compact presentation uncorrelated events
are built and weighted
 in proportion to real events of each multiplicity.
 For each bin in $Z_{tot}$, fixed at six atomic number units, an exponential
 evolution of the correlation function is observed from $\Delta$Z=7-8 down to
 $\Delta$Z=2-3. This exponential evolution is thus taken as ``background'' to
 extrapolate down to the first $\Delta$Z bin.
  Higher order  correlation  functions for the first bin in $\Delta$Z
  are displayed in
 Fig.~\ref{cor_exp_quant} with their statistical errors;
 the full line
 corresponds to the extrapolated ``background''. All events corresponding to
 the points whose error bar is fully located above this line correspond to
 a statistically 
 significant enhancement of equal-sized fragment partitions.
 The probabilities that these values higher than the background simply arise
 from statistical fluctuations are in the range 0.05-0.02
 depending on $Z_{tot}$. 
 The number of significant events amounts to 0.1\% of the selected fusion events.

\begin{figure}
\includegraphics[width= 14cm]
{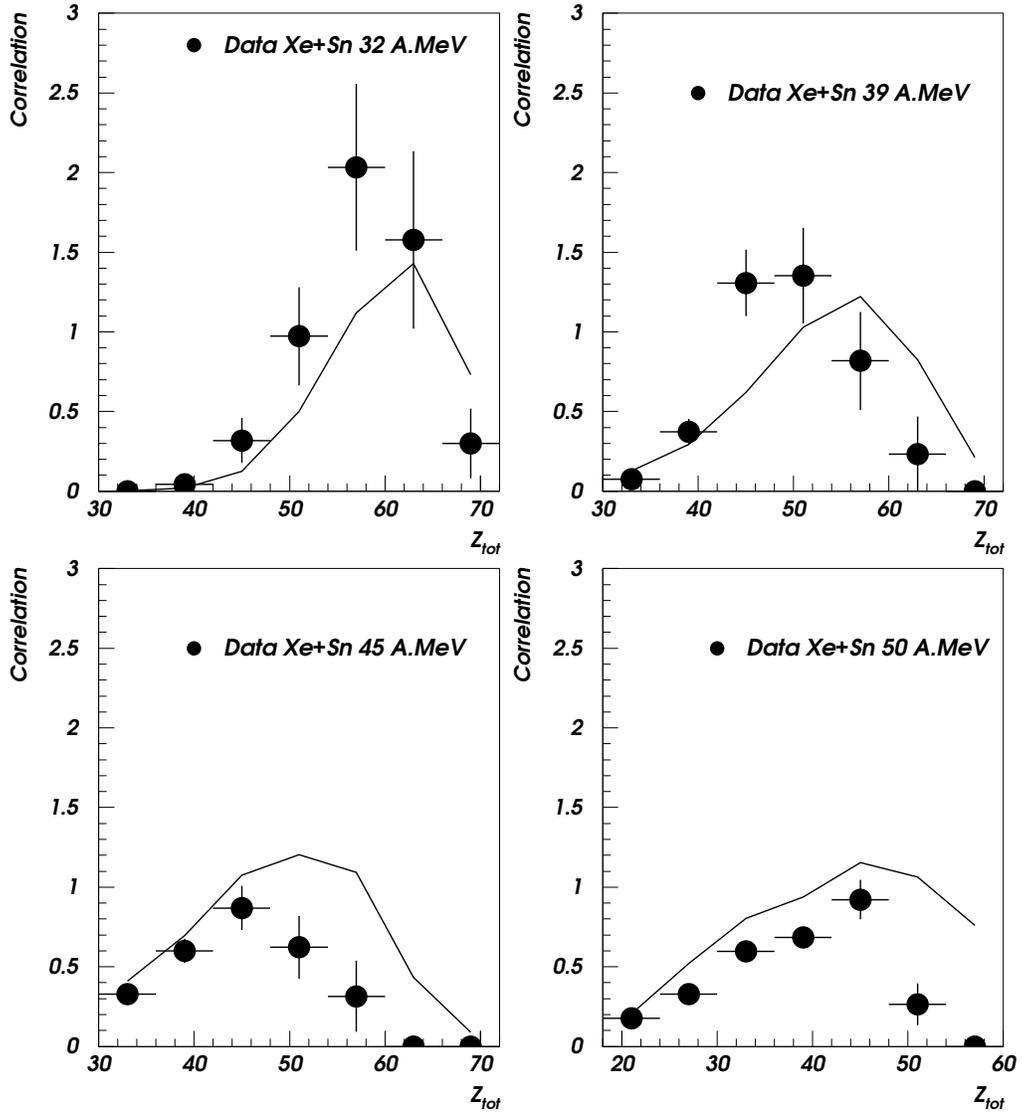}
\caption{Higher-order charge correlations: quantitative results for
experimental data at different incident energies.
Symbols indicate the events where $\Delta$Z = 0-1, curves show
the background (see text).
Vertical bars correspond to statistical errors and horizontal
bars define $Z_{tot}$ bins.}
 \label{cor_dife}
\end{figure}

This proportion has to be compared with what is expected for a complete 3D
simulation in which all events arise from spinodal decomposition. In this aim,
dynamical stochastic mean-field simulations~\cite{RA90,CHO91} were 
 performed for head-on collisions.
Spinodal decomposition was simulated using 
the Brownian one-Body (BoB) dynamics~\cite{CH94,GU97,FR002}, which consists in
employing a Brownian force in the kinetic equations. 
As a last step the spatial configuration of the primary fragments,
with their excitation energies as produced by BoB, was taken as input in the
SIMON code~\cite{ADN98} to follow the fragment deexcitation while preserving
space-time correlations. Finally the events were filtered
to account for the experimental device. These complete simulations
well reproduce multiplicity and charge distributions of fragments and 
their average kinetic energies~\cite{FR002}. 
Although all events in the simulation arise from spinodal decomposition,
the proportion statistically significant of equal-sized
fragment partitions is 
similar (0.15\%) to the experimental one.

A detailed quantitative comparison is
displayed in fig.~\ref{cor_exp_quant}. The similarities between experimental
and calculated events allow to attribute all fusion-multifragmentation
events to spinodal decomposition. The peaks observed near $\Delta Z$=0 
in the higher-order charge correlations are thus fossil
fingerprints of the partitions expected from spinodal decomposition.

What are now the quantitative experimental results for the different
incident energies above 32 AMeV ?
The energy dependence of the correlation function for fused events where
$\Delta$Z = 0-1 is shown in fig.~\ref{cor_dife}. No enhancement of nearly
equal-sized fragment partitions is observed for the two higher energies: 45
and 50AMeV. A similar negative result was mentioned in ref.~\cite{MO96} for
the Xe+Cu system at 50 AMeV. At 39 AMeV incident energy three points are
located above the extrapolated background. The probabilities that these
values simply arise from statistical fluctuations are 0.13 for $Z_{tot}$=39,
0.0002 for $Z_{tot}$=45 and 0.14 for $Z_{tot}$=51. The observed enhancement
amounts to about 0.2\% of events.

For the same selected fused events negative microcanonical heat capacities,
which are predicted to sign a first order phase transition~\cite{CHO99}, have
 also been
measured at 32 and 39 AMeV~\cite{LEN00}. The coincidence with present
observations is very appealing. Indeed, supported by theoretical simulations,
the observed weak but unambiguous enhanced productions of events with
equal-sized fragments at 32 and 39 AMeV can be interpreted as a signature of
spinodal
instabilities as the origin
of multifragmentation in the Fermi energy domain. Moreover the occurrence of
spinodal decomposition signs
the presence of a liquid-gas coexistence region and consequently, although
indirectly, a first order phase transition.

So dynamical (spinodal instabilities) and statistical
(negative heat capacities) arguments are in favor of a first order phase
transition. The following scenario can be proposed: spinodal instabilities
cause multifragmentation but when the system reaches the freeze-out stage, it has
explored enough of phase space in order to be describable through an
equilibrium approach.
The agreement between data and
dynamical and statistical models~\cite{FR002,BOU97,LEN99}
for static (multiplicity, Z distribution,
size of heaviest fragments) and kinetic properties of fragments fully
supports this scenario.

\section{Fragment charge distributions, power laws and event distributions}

\begin{figure}
\includegraphics[width= 14cm]
{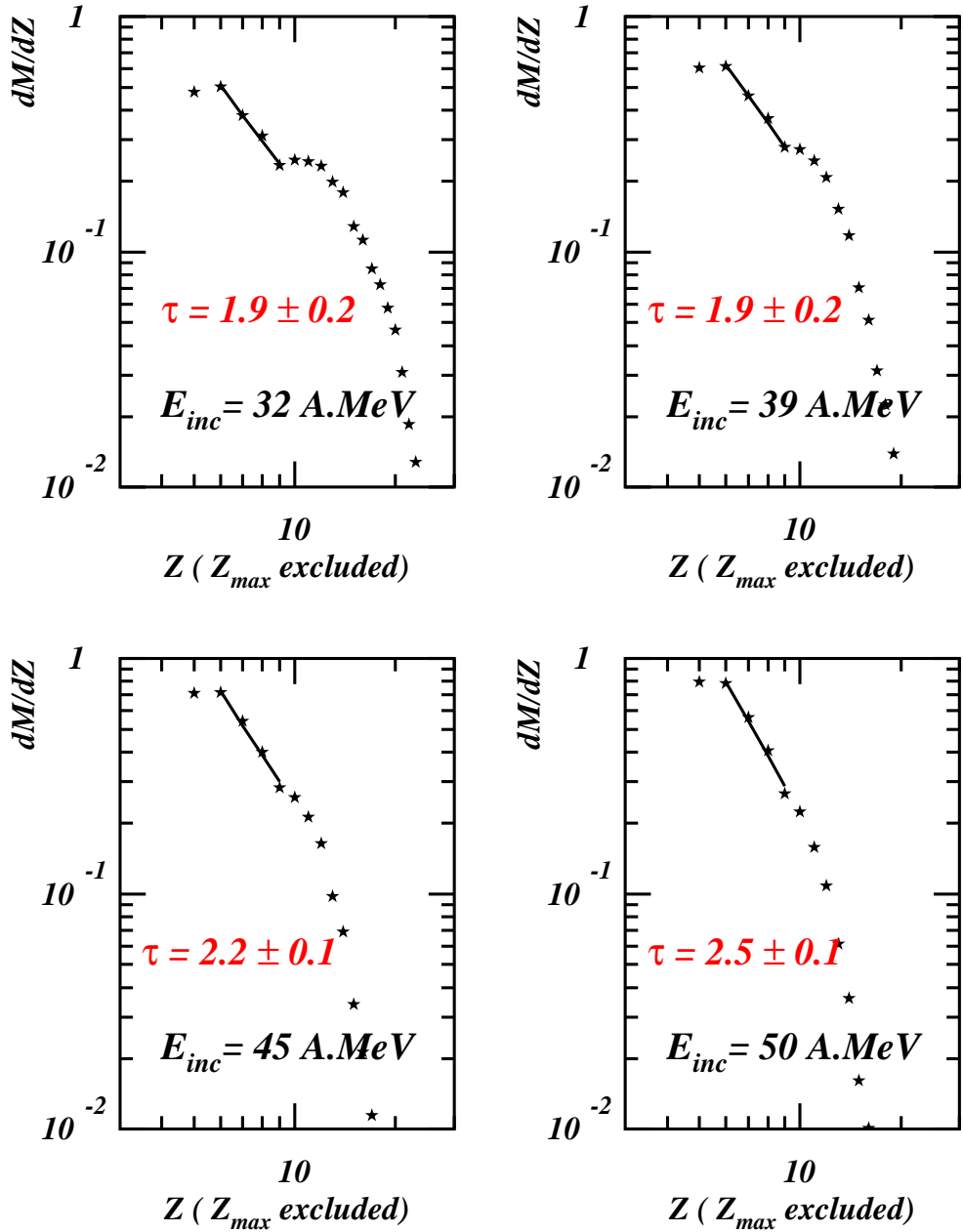}
\caption{Fragment charge distributions observed at different incident energies.
$\tau$ values refer to power laws observed (full lines).
 \label{tau_dife}}
\end{figure}

A power law behavior, which was claimed to be observed in the charge (mass)
distribution of multifragmenting systems, has been sometimes interpreted as
an evidence for a second order phase transition. Indeed, in the Fisher
droplet model a power law behavior is expected at the critical point when the
liquid and gas chemical potentials are equal and the surface tension is
zero~\cite{FI67}.
More recently a great theoretical effort was done, using Classical Molecular
Dynamics~\cite{PR95}, Lattice Gas Model~\cite{GU99,GU00} and statistical
models~\cite{RA97,RA01}, to give 
information on
fragment distributions when multifragmentation of
finite systems occurs in the coexistence region. Two of the main conclusions
are the following:

i)power laws are observed in the coexistence region which disappear as soon as
large systems are considered;

ii)within the microcanonical framework the
exponent of the power law is close to the one expected for the liquid-gas
universality class ($\tau$=2.33) only at the critical point.

Experimentally the well defined multifragmenting pieces of nuclear matter
presently studied can be used to bring some information.
Let us start considering the fragment charge distribution for the lower
incident energy (32 AMeV). At first glance a power law dependence with
$\tau$=1.1 rather well fits the Z distribution over the range Z=5-15.
Finite size effects
break this law for higher charges and the heaviest fragment is completely
excluded from this dependence. Removing this heaviest fragment we can
examine now in details the charge distributions measured at the different
incident energies. Such distributions are displayed in fig.~\ref{tau_dife} in a
double logarithmic scale. Valuable power laws are only observed on the reduced
 domain Z=6-9 and clearly a small bump with a maximum centered around Z=10-12
 is present in distributions for the two lower incident energies.
 $\tau$ values are equal to 1.9 for 32 and 39 AMeV incident energies. At
 higher incident energies for which neither negative microcanonical heat
 capacity nor fossil signatures of spinodal decomposition were observed,
$\tau$ values increase. Note that at 45 AMeV $\tau$ is equal to 2.2, i.e.
close to the one expected in the critical region.
Due to the reduced Z range where power laws are really observed, precise
experimental values of $\tau$ will be only obtained from fragment mass
distributions. 

\begin{figure}
\includegraphics[width= 14cm]
{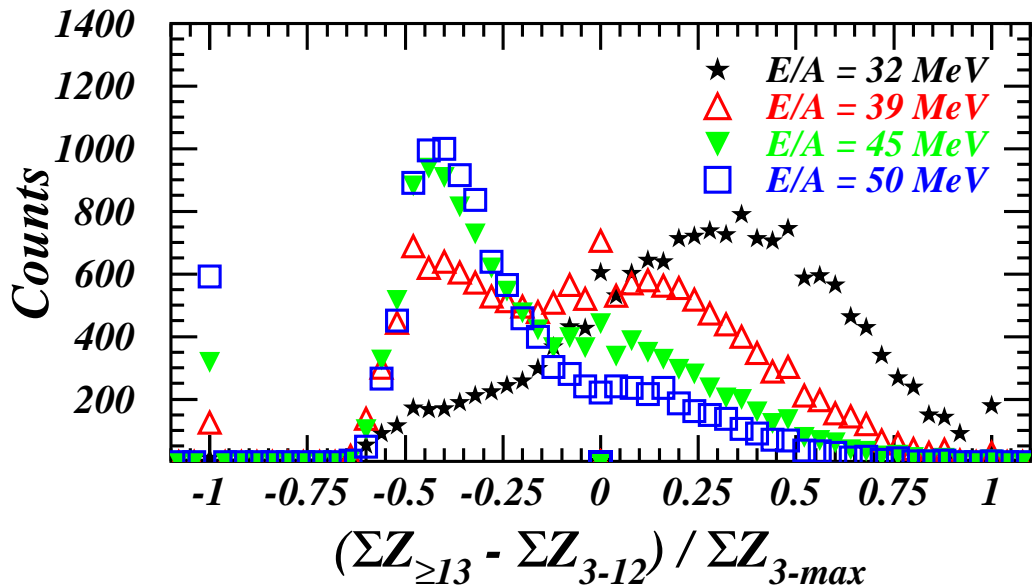}
\caption{Event distributions  observed at different incident energies. Points
located at the value -1 are divided by 10; See
text for the variable used for event distribution.
 \label{evdis_dife}}
\end{figure}

From the peculiar shapes of Z distributions observed in fig.~\ref{tau_dife}, it
is very tempting to consider that the bump centered around Z=10-12 can
indicate that other among the heavier fragments can belong to
 the liquid fraction; the heaviest fragment, generally
considered as the liquid fraction was already removed for distributions
presented in fig.~\ref{tau_dife}. With such an hypothesis power law concerns
only a part of the gas fraction.

We can try to go further by considering that a separation between liquid and gas
fractions is around Z=12. This separation, if not over-simple, can be
revealed in event distributions using as a variable the normalized
difference between the liquid and gas $Z_{bound}$. Fig.~\ref{evdis_dife} shows
event
distributions for the different incident energies considering only
Z$\geq$3; indeed light charged particles are a mixing of particles emitted at
different stages of the collisions and can not bring reliable information.
We observe a rather constant evolution of the distributions between 32 AMeV 
(dominated by the ``liquid fraction''), 39 AMeV (equilibrated ``mixed phase'')
and 45 AMeV (dominated by the ``gas fraction''). Inversely, distributions
at 45 and 50 AMeV are very similar and could characterize the fact that these
events are produced at the border or outside the coexistence region.
This part of the work is a first attempt to study new observables related to
the
existence of a first order phase transition. This is done in the spirit of
theoretical propositions made in ref.~\cite{CH01}. It is also discussed
in contributions~\cite{LO01,GU01} during this meeting.

\section{Summary}
In summary, the above studies illustrate:

$\bullet$an enhancement of nearly equal-sized fragment partitions in
multifragmentation of a heavy system formed on a restricted domain of
incident energies (32 and 39 AMeV);

$\bullet$that supported by dynamical calculations this enhancement is interpreted
as a signature of spinodal instabilities as the origin of multifragmentation
in the Fermi energy domain;

$\bullet$the coincidence, at 32 and 39 AMeV, between  observations of
spinodal instabilities and negative heat capacities;

$\bullet$that dynamical (spinodal instabilities) and statistical (negative
heat capacities) arguments are in favor of a first order phase transition in
finite systems ($\sim$200 nucleons);

$\bullet$that the following scenario for multifragmentation in the Fermi
energy domain can be proposed: spinodal instabilities cause multifragmentation
but when the system reaches the freeze-out stage, it has explored enough of
phase space to be describable through an equilibrium approach;

$\bullet$that, for fragment charge distributions, power laws are
only observed on a reduced domain in Z, from 6 to 9;

$\bullet$that $\tau$ is equal to 1.9 in the ``coexistence region''
and reaches a value close to the one expected in the critical region at
45 AMeV incident energy;

$\bullet$that the peculiar shapes of Z distributions observed in the
``coexistence region''  suggest that, as expected from spinodal decomposition,
more than one heavy fragment contribute to the liquid fraction;

$\bullet$that event distributions experimentally observed can also reveal
the existence of a first order phase transition.

\end{document}